\documentclass[sigconf,nonacm]{acmart}
\AtBeginDocument{%
  }

\usepackage{enumitem}

\begin{document}

% \acmYear{2026}\copyrightyear{2026}
% \setcopyright{cc}
% \setcctype[4.0]{by}
% \acmConference[FSE Companion '26]{34th ACM Joint European Software Engineering Conference and Symposium on the Foundations of Software Engineering}{July 5--9, 2026}{Montreal, QC, Canada}
% \acmBooktitle{34th ACM Joint European Software Engineering Conference and Symposium on the Foundations of Software Engineering (FSE Companion '26), July 5--9, 2026, Montreal, QC, Canada}
% \acmDOI{10.1145/3803437.3805535}
% \acmISBN{979-8-4007-2636-1/26/07}

\title{Test Before You Deploy: Governing Updates in the LLM Supply Chain}

\author{Mohd Sameen Chishti}
\affiliation{%
  \institution{Department of Computer Science \\ NTNU}
  \city{Trondheim}
  \country{Norway}
}
\email{mohd.s.chishti@ntnu.no}
\orcid{0000-0003-3977-8488}

\author{Damilare Peter Oyinloye}
\affiliation{%
  \institution{Department of Computer Science \\ NTNU}
  \city{Trondheim}
  \country{Norway}
}
\email{peter.d.oyinloye@ntnu.no}
\orcid{0000-0002-4925-5042}

\author{Jingyue Li}
\affiliation{%
  \institution{Department of Computer Science \\ NTNU}
  \city{Trondheim}
  \country{Norway}
}
\email{jingyue.li@ntnu.no}
\orcid{0000-0002-7958-391X}

% \affiliation{%

% \affiliation{%

% \affiliation{%

% \affiliation{%

\renewcommand{\shortauthors}{Chishti et al.}

\begin{abstract}
Large Language Models (LLMs) are increasingly used as core dependencies in software systems. However, the hosted LLM services evolve continuously through provider-side updates without explicit version changes. These silent updates can introduce behavioral drift, causing regressions in functionality, formatting, safety constraints, or other application-specific requirements. Existing approaches focus primarily on regression testing or versioning but do not provide deployer-side mechanisms for governing compatibility during opaque model evolution. This paper proposes a deployment-side governance framework based on three components: clearly defined rules for how the model is allowed to behave (production contracts), focused testing organized by deployment risk categories (risk-category-based testing suite), and release checkpoints that block updates unless they meet defined safety and performance standards (compatibility gates). Through exploratory validation across multiple LLM versions, we provide evidence that  targeted testing in specific risk areas can uncover performance regressions that overall metrics miss. We also identify several open research challenges, including how to systematically build effective test suites, how to set reliable performance thresholds in non-deterministic systems, and how to detect and explain model drift when providers offer limited transparency. Overall, we frame LLM update management as a software supply chain governance problem and outline a research agenda for putting deployer-side compatibility controls into practice.

%Through exploratory validation across multiple LLM versions, we demonstrate that slice-level evaluation can surface regressions that aggregate metrics fail to detect. We further identify key open research challenges, including systematic regression suite construction, threshold calibration under non-determinism, and drift attribution under limited provider transparency. Our work frames LLM update management as a software supply chain governance problem and outlines a research agenda for operationalizing deployer-side compatibility control.
\end{abstract}




\maketitle

\section{Introduction}

Large Language Models (LLMs) are increasingly embedded as core dependencies in software engineering processes, facilitating code generation, autonomous agents, and workflow automation pipelines \cite{10.1145/3586030}. However, unlike traditional software libraries, LLM services evolve much more rapidly. The LLM provider updates the model weights, safety policies, and serving infrastructure without changing the API endpoints \cite{OpenAI2023GPT4}. As a result, the downstream systems can experience drift in behavior, including reduced task accuracy, increased refusal of previously supported requests, and failures in structured or machine-consumable outputs. We refer to such unintended changes that can violate application expectations as behavioral drift. These silent infrastructure updates can introduce breaking changes in the production LLM-based systems. For example, Anthropic's August 2025 postmortem report documents three bugs affecting Claude Sonnet 4's performance without any API modifications, including routing errors affecting up to 16\% of requests and output corruption causing non-ASCII character injection \cite{AnthropicPostmortem}.  Similarly, \citet{Chen2024How} provide empirical evidence of drift by documenting that direct execution of code generated by GPT-4 drops from $52$\% to $10$\%  within three months (March-June, 2023), without any version change. This demonstrates how infrastructure changes can disrupt downstream applications despite stable endpoints. 

% provide  empirical evidence of drift, documenting that GPT-4's accuracy on a mathematical reasoning task dropped from 97.6% to 2.4% within three months, while code generation outputs introduced new formatting errors, both without any version change. 

% This demonstrates how infrastructure changes can disrupt downstream applications despite stable endpoints. Similar concerns have appeared in community forums\footnote{https://community.openai.com/t/months-of-curated-art-styles-broken-by-recent-image-model-changes/1370685} \footnote{https://support.google.com/gemini/thread/379487030?hl=en\&sjid=10232390588041134343-EU} across providers, although these reports remain largely unverified.

% Recent studies show that such silent updates can introduce performance regressions that destabilize deployed workflows and force developers to adapt prompts and system logic after deployment \cite{}. In practice, LLM providers typically communicate updates through high-level release notes or informal change logs, rather than managing them as governed dependency changes with explicit compatibility guarantees, regression disclosure, and audit-ready version artifacts.

These incidents highlight systematic challenges beyond individual provider failures. In traditional software supply chains, dependency changes are versioned, tested, and explicitly adopted \cite{cox2019surviving}, but hosted LLM services violate this assumption because their behavior may change without version or endpoint changes. We argue that behavioral drift represents a fundamental governance risk in the software supply chain. Unlike performance fluctuations, which are temporary and reversible, silent updates can permanently alter dependency behavior without deployers' control.
The provider-level evaluation also cannot anticipate application-specific compatibility requirements. Release notes describe general improvements but do not address specific constraints on formatting, error handling, or domain logic.
% not merely a performance fluctuation, as it undermines control, predictability, and accountability in dependency management.
% Behavioral compatibility is inherently application-specific; therefore, provider-level evaluations and informal update communications cannot reliably anticipate downstream impacts across diverse deployment contexts.

To manage LLM updates as controlled software dependencies rather than informal service changes, we propose a deployer-side governance framework with three components. First, production contracts define clear rules for how the model is allowed to behave (e.g., ``authentication code must pass security tests'' or ``JSON outputs must be valid''). Second, targeted tests focus on high-risk situations, such as security-sensitive tasks, format-critical outputs, or billing logic. Third, compatibility gates automatically check whether updates meet defined safety and performance standards and block deployment if violations occur, requiring review before a model update is adopted.

\section{Literature Review}

\citet{10.1145/3644815.3644950} demonstrate that silent provider updates frequently result in prompt performance regressions and recommend regression testing and version control, but they do not provide a concrete operationalization.
% forcing developers to adapt workflows post-deployment and exposing the limitations of classical regression testing assumptions for non-deterministic systems . It highlights that behavioral drift of LLM is not an anomaly but a recurring property of evolving services.
\citet{echterhoff-etal-2024-muscle} complement this work by studying the aftereffects of model evolution and showing that updates may introduce instance-level regressions despite aggregate performance improvements. They characterize the phenomenon in which newer model versions perform worse on specific tasks as negative flips. These findings indicate that global performance metrics cannot guarantee behavioral compatibility at the workload level.
% In order to improve reproducibility and change tracking in AI-enabled system, \citet{10.1145/3696630.3728714} proposed versioning framework for LLM-agent-based software artifacts. It tries to address the non-determinism and multi-artifact composition through behavioral fingerprints and extended semantic versioning schemes.  This approach is effective for governing developer controlled artifacts, yet, it may not address provider-driven behavioral drift in hosted LLM services where updates to models, policies and infrastructure occur outside deployer control.
Building on this work, \citet{10.1145/3696630.3728714} propose semantic versioning using behavioral fingerprints, which provide compact representations of model behavior across evaluation tasks to improve reproducibility. This approach is effective for developer-controlled artifacts, but it assumes version-control authority over the LLM itself, which may not hold when providers unilaterally update hosted models without changing endpoints. In practice, providers publish release notes but rarely provide behavioral compatibility versioning, regression disclosure, or machine-readable artifacts, forcing developers to infer update impacts reactively. Behavioral fingerprints capture aggregate drift but do not support application-specific and multidimensional compatibility requirements.

Beyond individual provider updates, broader work frames the LLM ecosystem as a complex software supply chain involving datasets, model hubs, tooling platforms, and centralized service providers. Studies \cite{10.1145/3746252.3761510, hepworth2024securing} highlight the risks involved in these AI pipelines, including opaque provenance, dependency vulnerabilities, and cascading failures. Industry-level security frameworks also identify LLM supply chain components as high-risk surfaces that lack transparency and control \cite{OWASP_LLM03_2025}. 

Prior work shows that silent updates can cause regressions \cite{10.1145/3644815.3644950,echterhoff-etal-2024-muscle} and proposes versioning for controlled artifacts \cite{10.1145/3696630.3728714}, but it does not operationalize deployer-side governance when providers update models unilaterally. Production monitoring tools such as Evidently AI\footnote{https://www.evidentlyai.com/} can detect drift in LLM output, but detection alone is not governance. A deployer still needs to determine which behavioral changes are relevant to their application and  how much deviation is tolerable. There should also be a policy on what should happen when drift exceeds the bounds. 
We address this gap by introducing risk-category-based regression suites that tie drift detection to application-specific deployment concerns, and compatibility gates that trigger structured review and mitigation workflows when contract thresholds are violated.

\section{Deployer-Side Update Governance Mechanism}
% We propose a deployer-side governance framework that treats hosted LLM services as continuously evolving software dependencies. It combines risk-category-based evaluation with explicit compatibility contracts and threshold-based adoption gates.

% We propose a deployer-side governance framework that treats hosted LLM services as continuously evolving software dependencies. Building on the requirement for regression testing established by \citet{10.1145/3644815.3644950} and addressing the findings of \citet{echterhoff-etal-2024-muscle} on instance-level regression, our framework combines  risk-category-based 
% slice 
% evaluation with explicit compatibility contracts. When drift exceeds application-defined thresholds, the compatibility gate triggers review and mitigation workflows, enabling controlled adoption of provider-side model changes. 

\subsection{Design Principles}

Non-deterministic service dependencies are not new to software engineering, and CI/CD quality gates have long been used to manage them. However, such services typically expose  versioning and operate under service level agreements. The hosted LLM  services are fundamentally different. They have natural language in the behavior space and minimal  version granularity.  The same prompt can produce different responses across runs. It shows that existing quality gate patterns cannot be applied directly, because the provider does not define a behavioral specification that deployer can test against. Our framework addresses this by shifting that responsibility to the deployer, who defines application-specific  contracts and gates updates these contracts.

Drawing on the gaps identified in prior works and nature of LLMs, the framework is guided by three principles that address issues  of regression testing, negative flips, and supply chain governance \cite{10.1145/3644815.3644950, echterhoff-etal-2024-muscle, OWASP_LLM03_2025}.

% This means that existing quality gate patterns cannot be applied directly, since there is no stable behavioral contract to gate against in the first place. It is guided by three principles:

% We propose a deployer-side governance framework that treats hosted LLM services as continuously evolving software dependencies. It combines risk-category-based evaluation with explicit compatibility contracts and threshold-based adoption gates.

\noindent \textit{Explicit compatibility contracts}:
Compatibility requirements must be defined as explicit behavioral rules with measurable thresholds, rather than inferred from provider benchmarks or aggregate performance metrics. This makes deployment conditions transparent, auditable, and tailored to application-specific risks.

\noindent \textit{Risk-centered evaluation}:
Compatibility must be assessed across the risk categories 
% slices 
 representing distinct  deployment concerns, rather than through a single global score. This prevents critical regressions, such as formatting failures, security vulnerabilities, or billing logic errors from being masked by overall performance improvements.

\noindent \textit{Deployer-controlled adoption gates}:
Model updates should not be adopted solely on the basis of provider-side releases or announcements. Instead, deployer-side gates should compare observed  category behavior within each risk category against  thresholds and determine whether the update can be safely integrated or requires mitigation.

\subsection{Framework Components}

These principles are manifested through three conceptual components of the proposed framework. Our focus is to shift the update governance from provider announcement to deployer-controlled compatibility verification.

\noindent \textit{Production Contracts}:  The production contracts define explicit behavioral requirements that must hold for safe deployment.   The contracts specify measurable thresholds, such as minimum correctness rates, formatting constraints, and security requirements.

\noindent \textit{Risk-Category-Based Regression Suite}: The risk-category-based regression suite organizes evaluation prompts into risk-aligned groups that represent distinct deployment concerns. Each  risk category is evaluated independently to detect domain-specific drift that aggregate metrics can mask. The suite also logs LLM snapshots, each with a model identifier and execution timestamp, to enable reproducibility and drift attribution.

\noindent \textit{Compatibility gates}: The compatibility gate compares category-level metrics against production contract thresholds and triggers review workflows when any violation occurs. It shifts the adoption decision from provider-controlled releases to deployer-controlled compatibility checks.

Here, we use a code-generation assistant for backend systems to illustrate how these components work together. The production contract specifies that database-intensive code must pass a defined set of unit tests and that all generated JSON outputs must conform to strict formatting rules. The risk-category-based regression suite organizes prompts into three categories: 
1) Category 1 contains prompts that generate complex backend logic with SQL interactions, execute the generated code against test cases, and compute a pass-rate metric; 2) Category 2 contains prompts that generate structured JSON outputs and validate formatting compliance; 3) Category 3 contains standard security audit tests for generated code. The testing suite records the snapshot of the LLM provider, such as ``gpt-4-turbo'' at timestamp 2026-01-26T10:00Z. The compatibility gate reviews these risk-category-level metrics against contract thresholds. It can be like Category 1 needs a pass rate of more than 95\% in addition to some mandatory tests. Category 2 requires 100\%  format compliance, and Category 3 requires 0\% security violations. 

% The framework is organized around three conceptual components, production contracts, slice-based regression suits, and compatibility gates. The production contract define the behavioral condition that must hold for safe deployment such as minimum correctness or formatting thresholds. The slice-based regression suites operationalize these contracts  by grouping evaluation prompts according to specific risk areas. The suite also calculate the slice-level metrics such as pass rate or compliance scores. The compatibility gates interpret these metrics against predefined thresholds and trigger review or mitigation workflows when drift is detected. Together, these components can shift the update decision from LLM provider announcements to deployer-controlled compatibility checks.

% For example, in a code generation assistant for backend systems, a production contract may specify that database-intensive code must pass a defined set if unit tests and that all generated JSON outputs must follow strict formatting rules. Slice 1 contain the prompts that generate complex backend logic with SQL interactions, execute the generated code against test case and compute a pass-rate metric. Slice 2 contain prompt that generates a structured JSON output and check if formatting. Another slice or test suits may contain standard test for security auditing of the generated test code. The compatibility gate then review these slice-level metrics against thresholds and triggers review if any required test, formatting condition or security audit is violated.

\subsection{Operational workflow for managing LLM behavioral drift}

The operational workflow translates these components into a systematic process for governing LLM updates.

\begin{enumerate}[leftmargin=*]
    \item Specify production contracts that define unacceptable breakage for each application context.

    \item Organize benchmark items into risk-aligned categories , with each item executed multiple times to produce stable metrics, and record results alongside prompt and model versions.

    \item Compute risk category-level metrics and compare against contract thresholds.
   
    \item Trigger mitigation actions when drift exceeds thresholds (prompt adaptation, workflow revision, fallback activation, and revalidation).

\end{enumerate}

\begin{figure}
    \centering
    \includegraphics[scale=0.25]{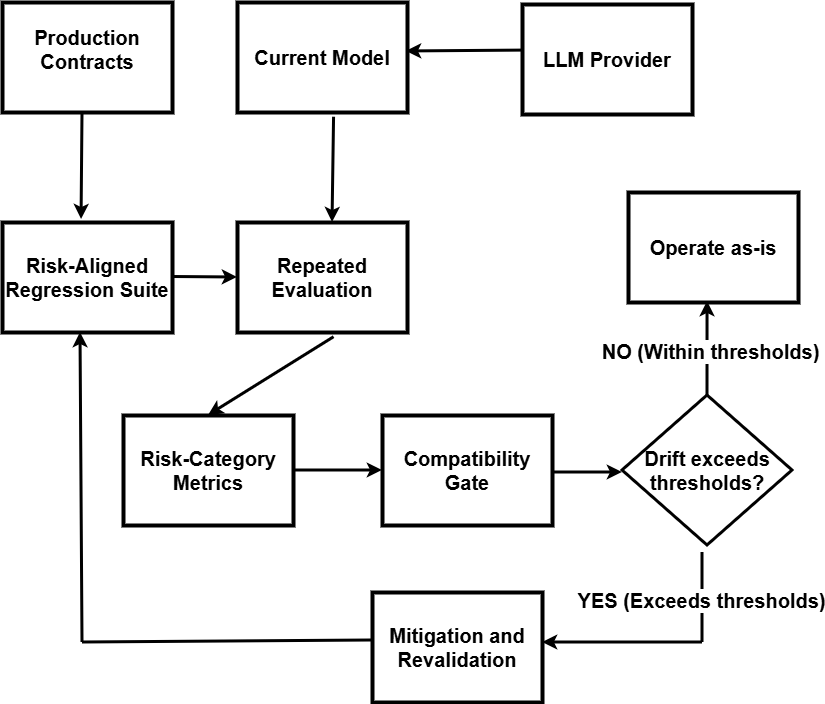}
    \caption{ Deployer-Side LLM Drift Governance Workflow.     }
    \label{fig:regframe}
\end{figure}

Figure \ref{fig:regframe} visualizes the deployer-side governance workflow, showing how production contracts are translated into risk category-based regression suites, evaluated across evolving LLM snapshots, and used to trigger drift detection, mitigation cycles. These steps produce contextual regression suites and continuous compatibility evidence, enabling organizations to govern LLM dependency updates through compatibility gates that trigger review before adoption.

\section{Preliminary Validation}

We tested the framework to see if it works in practice and what challenges arise during implementation. This is an exploratory validation, not a comprehensive evaluation. We wanted to verify whether the organization can define production contracts, build risk-category-based test suites, and implement compatibility gates to detect drift.

\subsection{Experimental Setup}
We evaluated behavioral drift across model transitions using Anthropic’s Claude models via the web interface (Haiku 3.5, Opus 3, Sonnet 4, Opus 4.5, Sonnet 4.5, Haiku 4.5, and Opus 4.6) to represent a range of model sizes and performance tiers, which enable us to observe drift patters across incremental updates (Sonnet 4 to 4.5) and tier transitions (Haiku to Opus). These models were chosen because they represent complete Anthropic's product line and  its documented incident provided a known case of silent drift for validation. As the web interface does not expose immutable version identifiers or deterministic configuration controls, snapshot tracking remained approximate and relied on visible model names and timestamps.

We designed 25 prompts across three task domains \cite{llm_drift_supplementary_2026}, which we treated as distinct categories aligned with deployment risks:

\begin{itemize}[leftmargin=*]
    \item Authentication functions (password hashing, JWT validation, session management)
    \item Data validation (email validation, SQL injection prevention, input sanitization)
    \item Structured output generation (JSON formatting, CSV output, log formatting)
\end{itemize}

Each prompt specified explicit behavioral requirements, which were treated as production contract conditions. These included:
\begin{itemize}[leftmargin=*]
    \item Functional correctness (For example, code must pass predefined test cases)
    \item Constraint adherence (For example, strict JSON format compliance)
    \item Instruction compliance (For example, ``code-only'' output without explanation)
\end{itemize}

For each model configuration, responses were generated manually. Function code was extracted and evaluated against predefined criteria. Functional correctness was tested using automated execution where feasible, while formatting and instruction compliance were assessed through structured inspection.
Each prompt was executed multiple times (3–5 runs) to observe output variability and estimate risk category-level compliance behavior.

 % \vspace{-0.3cm}
\subsection{Findings}

The exploratory validation revealed several forms of behavioral drift consistent with the risks targeted by the framework. We observed regressions in format compliance across model transitions. Sonnet 4.5 returned premature errors on JSON tasks, while Haiku 4.5 and Opus 4.5 changed exception types. Sonnet 4, 4.5 and Opus 4.6 returned empty JSON due to  security concerns while Haiku 3.5 generated the JSON for prompt with missing data. Haiku 3.5 and Sonnet  also added explanatory text despite code-only requirement. We also observed the risk category-specific regressions in some prompts. The SQL and authentication functions remained stable, while structured JSON tasks showed higher drift. Sonnet 4.5 and Opus 4.6 failed  item-processing tasks, Sonnet 4  generated JavaScript instead of Python and Opus 4.5 wrapped outputs in metadata objects.  Moreover, a backend SQL function passed all tests with Sonnet 4, but failed a test for safe encoding next day, suggesting silent infrastructure changes. Sonnet 4.5 output remained stable for both days.

Overall, while the validation was limited in scale and manual, it shows that production contracts and risk-category-based evaluation can surface regressions that are not visible through aggregate correctness metrics alone.

% \vspace{-1cm}

\section{Limitation and Open Challenges}

Our exploratory validation exposed structural gaps that limit the immediate operationalization of deployer-side drift governance.  These gaps define the research agenda for systematic compatibility governance. Our framework provides the mechanism, however, the challenge lies in constructing appropriate test suites. Our regression prompts were manually constructed based on anticipated failure modes. While this approach surfaced meaningful regressions, it lacks principled coverage guarantees. Unlike traditional software testing, where techniques such as equivalence partitioning, boundary value analysis, and mutation testing provide structured coverage criteria, LLM evaluation operates on natural-language inputs and implicit behavioral expectations. A fundamental open question here is, \emph{How can we design regression suites that achieve meaningful coverage of deployment-relevant behavior in LLM systems?}
We need further research to define coverage metrics for prompt-based systems, including notions of behavioral diversity and risk-weighted sampling across risk categories, among others.

Our experiments relied on intuitive thresholds for pass rates, formatting compliance, and other parameters without principled statistical calibration. However, strict thresholds may generate excessive false alarms, while lenient thresholds may mask critical regressions. It raises two interrelated challenges, how should risk category-level thresholds be calibrated based on deployment risk and how can compatibility decisions incorporate statistical confidence in non-deterministic systems?
% Future work must explore these aspects.

% confidence intervals for behavioral stability and risk-adjusted gating strategies aligned with application criticality.

Our repeated executions revealed variability in compliance outcomes for identical prompts. Some responses passed all checks in one run and failed in another. This complicates binary pass/fail judgments and raises deeper questions about what constitutes stable compatibility in stochastic systems. We need to determine how many executions are sufficient to establish behavioral stability. We also need to find which statistical models best capture risk category-level drift under non-determinism.

% 
% % We also observed that behavioral changes occur both when switching between model versions and when using the same model on different days.
% 
% However,  LLM providers
% % most LLM providers do not clearly disclose what exactly changed in the system. 
%  rarely give detailed update logs or information about infrastructure modifications. 
Continuous category-level evaluation introduces computational overhead and operational complexity to CI/CD pipelines. Organizations must balance regression testing costs against deployment velocity and risk exposure. We need to find new adaptive testing strategies that reduce costs while preserving risk coverage. We also need to look for governance patterns that are sustainable for production LLM deployments.

Lastly, hosted LLM providers rarely expose version identifiers or fine-grained update disclosures. his makes it difficult to understand the cause of observed drift. A change in behavior could result from updated model weights, modified safety policies, changes in the serving system, or simply normal randomness in generation.
Without provider-side transparency, deployer-side governance operates under partial observability.

% \vspace{-0.2cm}

\section{Conclusion}

Hosted LLM services introduce a new form of software dependency in which behavioral changes may occur without explicit version control or transparency. This undermines traditional assumptions of dependency management and creates governance risks for production systems. We proposed a deployer-side compatibility framework grounded in explicit production contracts, risk-aligned category evaluation, and threshold-based compatibility gates. Our exploratory validation illustrates that risk category-level analysis can reveal regressions masked by aggregate metrics and that silent drift may occur even without formal model version transitions. While the framework is feasible with existing tooling, its operationalization raises important open challenges in regression suite design, non-deterministic evaluation, threshold calibration, drift attribution, and computational overhead. Addressing these challenges is a necessary step towards 
% transforming LLM update management into a 
principled supply-chain governance practice. 
% By reframing LLM evolution as a compatibility governance problem, this work lays the foundation for systematic deployer-controlled update validation in the LLM supply chain.

\bibliographystyle{ACM-Reference-Format}
\bibliography{sample-base}


\end{document}